\documentclass[12pt,letterpaper]{article}
\usepackage[a4paper, total={7in, 10in}]{geometry}

\usepackage{graphicx}
\usepackage{helvet}
\usepackage{authblk}
\usepackage{hyperref}
\usepackage{amsmath} 
\usepackage{amssymb} 
\usepackage{orcidlink} 
\usepackage[super,comma,sort&compress]  
   {natbib}
\usepackage[right]{lineno} \linenumbers

\makeatletter
\renewcommand{\maketitle}{\bgroup\setlength{\parindent}{0pt}
\begin{flushleft}
  \textbf{\@title}
  
  \@author
\end{flushleft}\egroup}
\makeatother

\title{A Field Biology Guide for the Curious Physicist}
\date{}

\author[1,*]{S. David Stupski}
\author[2, *]{Laura Casas Ferrer}
\author[2, *]{Jacob S. Harrison}
\author[3] {Justina Jackson}
\author[4]{Carolina Paucarhuanca Mansilla}
\author[4] {Loribeth Maricielo Bolo Lívano}
\author[5] {Avaneesh Narla}
\author[6] {Chew Chai}
\author[7] {Elizabeth Clark}
\author[2] {Nami Ha}
\author[8] {Jaime Quispe Nina}
\author[9] {Ethan Wold}
\author[10]{Johana Reyes-Quinteros}
\author[4,10,11]{Geoffrey Gallice}
\author[2]{Saad Bhamla}

\affil[*]{Authors contributed equally}
\affil[1]{Department of Biology, University of Washington, Seattle, WA, USA}
\affil[2]{School of Chemical and Biomolecular Engineering, Georgia Institute of Technology, Atlanta, GA, USA}
\affil[3]{Center for Education Integrating Science, Mathematics, and Computing, Georgia Institute of Technology, Atlanta, GA, USA}
\affil[4]{Department of Engineering, Pontifical Catholic University of Peru, Lima, Peru}
\affil[5]{Department of Applied Physics, Stanford University, Stanford, CA, USA}
\affil[6]{Department of Bioengineering, Stanford University, Stanford, CA, USA}
\affil[7]{Advanced Light Source, Lawrence Berkeley National Laboratory, Berkeley, California, USA}
\affil[8]{Nature Travel Photography, Tambopata, Madre de Dios, Perú}
\affil[9]{School of Biological Sciences, Georgia Institute of Technology, Atlanta, GA, USA}
\affil[10]{Alliance for a Sustainable Amazon, Potomac, MD, USA}
\affil[11]{Department of Natural History, Florida Museum of Natural History, University of Florida, Gainesville, FL, USA}

\affil[*]{Correspondence: saadb@chbe.gatech.edu}

\begin{document}

\maketitle

\section*{SUMMARY}

Fieldwork is an essential component for an expanding umbrella of research on the physics of living systems, where observing organisms in nature is a critical component of discovery. However, conducting field research can be a barrier for scientists, in particular physicists, who do not have experience working with organisms under challenging field conditions. Here, we propose seven critical steps for organizing and executing interdisciplinary, curiosity-driven field research. Our recommendations are drawn from insights gained from the \textit{in situ} Jungle Biomechanics Lab, a field research course that helps early-career scientists from both physical and life sciences gain experience in both organizing and conducting interdisciplinary field research in the Amazon Rainforest. We emphasize a curiosity-driven approach towards the scientific inquiry of living systems, which we believe is crucial for discovery while working with wild organisms under unpredictable field conditions. We further provide guidance on teamwork when conducting fieldwork, including creating an inclusive environment and advocating for codes of conduct and team structures that aid in conflict resolution. Finally, we outline an in-situ approach to fieldwork that requires engagement with the environment, scientific community, and local people where field sites exist.

\section*{KEYWORDS}

field research, curiosity-driven science, biodiversity, scientific education, in situ studies, environmental identity
\section*{INTRODUCTION}

Whether it is the superpropulsion of planthopper pee droplets \cite{challitanaturecomm2023}, information propagation in swarms of fireflies \cite{sarfati2021self}, or a caterpillar's ability to detect predators' electric fields \cite{england2024prey}, many organisms are operating at the limits of what is physically possible. This has led to an increased interest in physical scientists and engineers utilizing their training in physics, mechanics, and instrumentation to help them understand the physics of living systems.  However, those organisms are not always amenable to being brought into a controlled laboratory environment, may only perform extreme behaviors in the wild, and often exist in biodiversity hotspots in relatively remote places- often requiring scientists to meet organisms where they live.  

Studying organisms in the field presents an array of challenges that can initially be daunting not just from a scientific perspective but also from logistical and conceptual standpoints.  What are the expected costs of fieldwork? What happens if I cannot find the organism that I set out to study? What challenges does field work present in team leadership?  How do I even find a research question to begin with?  What we aim to do here is provide a field guide for scientists who have not had the opportunity to begin field research with wild living organisms in their natural settings. This guide is woven together from lessons learned after running the Jungle Biomechanics Lab, an NSF-IRES program that takes scientists from expertise ranging from fluid mechanics to theoretical physics into the Peruvian Amazon. It draws on insights from scientists who have been doing fieldwork with living organisms for more than a decade, but critically also includes the perspective of physical scientists doing biological fieldwork for the first time.  We will approach field research from a curiosity-driven science perspective, which we believe is amenable to the uncertainty that field research often presents. Finally, we will advocate for what we call the \textit{in situ} approach to fieldwork, an approach that we think of as antithetical to parachute science and requires engagement at the level of the environment, the scientific community and the greater community where field sites exist.

\section*{Finding a field site}  
  A field site may be a literal field, a suburban backyard, or even the sewers of New York City \citep{guo2023systematic}.  Anywhere organisms have adapted to survive, there is likely unexplored biophysics to discover. However, the more challenging the environment, the more difficult it might be for a scientist to gain access and make observations. Identifying the perfect place to begin studying a system in the field requires a careful balance of acquiring funding and choosing field sites that have the infrastructure that can support your science. We suggest starting with a budget. This can determine whether or not you should begin your field research in your backyard or in the depths of the Amazon rainforest, both of which can yield important scientific insights for organismal biophysics.
  
With a budget in mind, it is time to decide on \textit{where} to go. Finding a place to do field work may be as simple as using Google Maps to look for freshwater ponds, in the case where one is setting out to study dogfighting dragonflies. For questions that can only be answered with organisms further from reach, biological field stations can be a critical resource for conducting field-based research with living systems.  Biological field stations offer infrastructure that might be necessary for executing prolonged and technically challenging research. The resources at field stations are often on how established or remote the station is, and the fees are often correlated with the number of benefits that can be available. Some stations may have full wet-lab facilities and dormitories, whereas others may have limited electrical access.   There are thousands of biological field stations globally, and databases exist to help locate a suitable research location for conducting biological field research  \citep{tydecks2016biological}. 

So, what can you do if field research is inaccessible to you? First, many field stations offer opportunities for research assistants, where room and board are often covered in return for assisting with ongoing research. Many universities and institutions offer travel grants and scholarships specifically for international field research. Check with your institution's financial aid office or research department for available funding opportunities. Organizations like the Society for Integrative and Comparative Biology (SICB), the Company of Biologists, and the Ecological Society of America (ESA) provide grants and fellowships for students and early-career researchers to engage in international research. Faculty can incorporate field research in their grants. The genesis of our own Jungle Biomechanics Lab was as part of the Broader Impacts of S.B.'s NSF CAREER proposal~\cite{noauthor_undated-ft}, which later evolved into a full NSF grant through the IRES program~\cite{noauthor_undated-ci}. Foundations such as National Geographic, Moore, and Fulbright also offer grant opportunities for education-related activities.  

\section*{Finding Research Questions In Organismal Biophysics: A Curiosity-Driven Perspective}
The modern academic science workflow pressures scientists to publish and to publish quickly. To a behavioral ecologist, identifying lines of scientific research can often feel similar to the ``explore/exploit" problems that animals face.  Just like an organism needs to balance its time between using the resources readily available (exploit) and looking for new resources for when current supplies dry up (explore), scientists must also balance time between collecting data for lines of questions that are producing publishable results and finding new lines of questions to build a career on. For many career scientists, it can then be difficult to dedicate time purely towards the ``explore" process. Developing new lines of research in organismal biophysics, however, demands slowing down- and taking the time to simply observe. 

An approach we advocate is to go into the ``explore" phase with a curiosity-forward strategy.  A curiosity-driven approach emphasizes observation as the motive force behind discovery.  Many significant findings in organismal biophysics begin with simple observations of organisms-more often than not by accident.  To name just a few: squirrel acrobatics \citep{science2021hunt}, dog slurping \citep{gart2015dogs}, springtail aerodynamics \citep{PNAS2022ortegajimenez}, cellular origami \citep{flaum2024curved}, cat grooming \citep{noel2018cats}, insect excretion \citep{challita2024unifying}, caterpillar electro-sensation \citep{england2024prey}, planthopper gears \cite{burrows2013interacting}, snapping shrimp \citep{patek2004deadly}, firefly flashes \citep{sarfati2021self}, elephant trunks \cite{schulz2022skin} and leaping eels \citep{catania2016leaping}. We see the curiosity-driven approach to "exploration" in action in the Jungle Biomechanics Lab. The JBL is a natural experiment that asks what happens when we bring early career researchers from fields ranging from theoretical physics to microbial ecology into an unfamiliar field site with high levels of biodiversity. The course is designed with structured spontaneity at its core, giving early career researchers unfettered room to spend time solely in an ``explore" phase.  What happens is that under these conditions, researchers stop, think, and observe, and without exception, quickly find an unresolved research question: whether in the flight control of cicadas who have lost their entire abdomen due to a \textit{Massaspora} infection or when ant lion larvae choose to build their sandy pits. A curiosity-driven approach requires open-mindedness to be flexible and not take seemingly simple behaviors that are physical marvels for granted. Curiosity, however, is something that is necessary to practice. As George M. Whitesides~\cite{Whitesides2018-wb} aptly observed,
\begin{quote}
    'Because following curiosity can seem effortless, it is easy to assume it does not need to be learned, practiced, or encouraged, that it is not important, and that it will somehow take care of itself. But, as with many activities that are competing for time and attention in a utilitarian world, curiosity can atrophy from neglect. It can certainly be unfocused, and lead to nothing (or at least nothing immediately useful), but using it as the starting point for careful observation of nature and society is a nontrivial skill, and a starting point for new intellectual endeavors and adventures. It is one essential contributor to creativity in science, and a start in forcing new ideas into inflexible professional orthodoxies.'

\end{quote} 
 
\section*{Preparing for field work: safety before science}
Preparation for any fieldwork can vary vastly depending on where the field site is, the study organisms, the experience level of your research team, and what equipment is needed to answer any scientific questions.  Preparing for fieldwork an hour's drive away from your home or university may require a different approach than organizing field research on another continent. Therefore, it is difficult to give all-encompassing guidance on exactly how to prepare before embarking on your specific project.  Instead, we offer a simple frame of mind for the early planning stages that have provided scaffolding in our own preparation, which is to place safety and transportation as first order priorities.  

Safety and transportation logistics for field work should always begin with a detailed risk management plan. Your risk management plan should include an outline of basic expectations for safe conduct and operating procedures for an emergency. As a reference, we have included one for the Jungle Biomechanics Lab as a supplement (S1). Critically, while preparing any safety plan for field research, you should be cognizant that safety risks can vary among individuals from different backgrounds and identities\citep{demery2021natureec}. 

In addition to a safety plan while in the field, a strong fieldwork plan lays out detailed logistics regarding travel. Local field work may only need a carpool sign-up sheet to ensure equitable access among participants. Organizing and leading field work farther away than a short drive is significantly more complicated and can be more constrained by a budget.  It is also imperative that you understand the local laws regarding collection and conducting research wherever your field site may be. Different countries may have specific laws about exporting samples if you intend to return with preserved specimens. Although seemingly obvious at first, specific details around travel logistics can get lost in early preparation steps, and we recommend having an explicit travel plan that outlines how each participant will get to the field site, which elements each person is responsible for, and a backup plan in case something should impede travel.  A strategy which builds a detailed logistical foundation will facilitate more technical and specialized research. However, we would like to emphasize that many great discoveries in organismal biophysics required only a keen eye and no specialized equipment at all. Much of our own research is implemented with phone cameras and some extra lighting. As a reference, we include example gear lists for our own field work. Our mindset is that a successful round of field work is one that gets all participants on site and home safely, and a tremendously successful field excursion is one in which participants return energized with ideas, data, and often a new research direction. 

\section*{Establishing a culture of mutual respect}
Collecting data in 40\textdegree~C and high humidity for multiple days in a row is hard. It can be even harder if the internet connection is limited, you have forgotten a crucial piece of equipment for your experiments, or you have not been able to find any of your species in the field. In the moment, miscommunications occur, and interpersonal conflict is inevitable. Disagreements, of course, are not signs of dysfunction, but how disagreements are handled is crucial for fostering a creative research environment.  A code of conduct, as well as a clear leadership structure, proactively outlines expectations for participants well before the field work begins, provides a clear throughline for conflict resolution, and helps tackle most interpersonal issues before they arise.

A code of conduct is a document that explicitly states the rules and values of a team or organization and includes clear consequences for violating such rules. For bringing scientists into the field for  we recommend using the Field Code of Conduct developed by the Association of Polar Early Career Scientists (APECS) to establish expectations among researchers \citep{apecs_field_code_conduct}.  The APECS code of conduct has clear definitions for what constitutes harassment and a zero-tolerance policy for harassment, as well as an internal leadership structure that provides both on-site and off-site independent contacts to report and initiate actionable consequences for inappropriate behavior.  All participants should read and signed the code of conduct before beginning any fieldwork. The values outlined in the document included but were not limited to, not putting oneself or others in dangerous situations, respecting personal boundaries, and showing respect for the land, its organisms, and the broader community. Incorporating principles of respect and inclusion is imperative to empower the next generation of field scientists.

\section*{Collaborate Across Disciplines and especially with Naturalists}

Field stations are crossroads for wide-ranging corners of the scientific community from all career stages and disciplines.  The same field site may simultaneously be a temporary home to industrial engineers, ornithologists, or disease ecologists across career stages ranging from high school students to senior faculty levels. The confluence of expertise at field stations is an invaluable opportunity for unexpected collaboration.  Oftentimes, executing complex field experiments requires input from expertise outside of one’s own knowledge domain or simply a few more sets of hands to perform experiments.  A key consideration when planning field research will always be to work with scientists from a broad range of expertise (for instance, a physicist interested in the locomotion of tiny organisms may benefit from the expertise of a cell biologist who has spent a career imaging organisms under a microscope).  Researchers with expertise in mechatronics may be indispensable for the \textit{ad hoc} fabrication of a force transducer from spare strain gauges, and a theoretical physicist may have a perspective for modeling the collective behavior of gregarious caterpillars. One additional collaboration we especially implore for the physicist beginning field biology is that with naturalists.

Naturalists are scientists who have expertise gained from years of careful observation and insights of organisms in their natural habitats and will be crucial for planning the practical elements of proposed field work, such as \textit{when/where can I even find my study organism?} Naturalists have a wealth of knowledge of organisms that is often difficult to extract from the literature alone. This knowledge has proven indispensable time and time again as the starting point for the physics-based inquiry of an organism's behavior.  For example, a naturalist might note that a family of butterflies, the metal marks (\textit{Riodinidae}), have a habit of landing upside down on the underside of leaves. At first, a seemingly simple and innocuous observation. However, for an expert in flight control, this can be understood as an inordinately complex task and has the potential to generate an entire line of research questions to dissect the control mechanisms of upside down take off and landing. The knowledge of naturalists is indispensable for identifying new lines of research questions that are suitable for the quantitative and experimental skills of physicists and engineers. Collaboration with naturalists should be a key component in any plans to do organismal fieldwork, especially at an unfamiliar location.

\section*{The \textit{in situ} approach to field biology is engagement on three levels}
In addition to finding field sites, organisms, and research questions, we especially want to emphasize what we describe as the \textit{in situ} mindset for fieldwork.  Here, we use the term \textit{in situ} to mean engaging both physically and mentally with the environment where field research takes place.  We intend for the \textit{in situ} approach to be an antithesis to parachute science, a practice where researchers from high-resource institutions drop in, extract samples, and leave without acknowledging or contributing towards local knowledge\cite{odeny2022time}. To us, the \textit{in situ} approach requires presence and engagement at three interlocking levels: the environment, the scientific community, and also the greater community where a field site exists.

\textit{In situ} engagement with the environment signifies understanding field research through the lens of an environmental identity. An environmental identity is something that each person has, whether they know it or not, and is a term used to refer to how we see ourselves in relation to the natural world \cite{clayton2003}. For example, a scientist's environmental identity may relate to the rainforest as a biodiversity hotspot with a new parasitic wasp waiting to be discovered.  For a subsistence farmer in the same region, the relationship between the rainforest and the person may be more about a day-to-day livelihood. Fieldwork can and should be understood in the social context of the environments in which it occurs.  \textit{In situ} within the local scientific community reflects understanding, that as scientists, we have a responsibility as mentors and colleagues. Scientists at all levels from localities where fieldwork occurs are invaluable collaborators and wherever possible should be included. For our own field research, this includes collaborations with scientists from local universities, and especially including undergraduates as active participants in research projects. Finally, \textit{in situ} at the community level reflects a commitment to being active participants not just in science but also in the local communities where the science takes place. Community involvement is an opportunity, a responsibility, and a privilege when conducting field research.  This may be performing outreach events showcasing research to grade school children, inclusion of local teachers and farmers or even establishing community scientific resources at field sites.  No single stratum of the three layers of an \textit{in situ} approach exists alone, and each is an indispensable for conducting field research.

\section*{Iterating Improvements from Quantitative and Qualitative Information}
Science is by nature an iterative process, and how we conduct and lead our scientific research is not less subject to continuous fine-tuning.  Both quantitative and qualitative data from field researchers are crucial in improving and iterating upon any field work led by a lab. Each, however, presents its own challenges to collect in a meaningful way. Taking an apocryphal queue from Galileo, we think of this as having two distinct halves:``measuring what is measurable" and ``making measurable what is not so". Getting data from participants before and after any complex field research is the first step in measuring the measurable. One potential tool is to use anonymous surveys. This way, you are more likely to have meaningful data that will allow you to improve any element of your field research that did not work as intended and can be immediately addressed, as an example we have included a survey we have used in the past as a supplement.  Qualitative data requires making measurable what is not so. Part of our philosophy was to give equal footing to all field researchers, regardless of whether they were postdocs or undergraduates. Within a culture of mutual respect, researchers, whether postdoc or undergraduate can freely express and suggest improvements that will lead to the completion of a project. Both elements are crucial for establishing a long-term succesful research program that includes challenging field work components. 

\section*{Concluding Remarks}
 In an interdisciplinary scientific community, organismal biology is enriched from the perspectives of physicists and engineers whose expertise in quantification, instrumentation, and computational frameworks is primed to answer fundamental questions about how organisms work.  What we hope to do here is build a larger table for scientists from more disparate fields to turn their efforts to biological systems which often requires us to do research where organisms have evolved to thrive. Field research is an integral part of organismal biology research. As more disciplines seek to draw inspiration from biological systems, it is critical that those fields engage the ecological context of their biological study systems. Starting a field biology project \textit{de novo} may seem overwhelming to those who have not had the opportunity, but we argue that it is worth the trouble. We see only opportunity in embracing more physical scientists into the naturalist traditions of the past few centuries and envision only expanding opportunities in the physics of living systems.

\newpage


\subsection*{Lead contact}


Requests for further information and resources should be directed to and will be fulfilled by the lead contact, Saad Bhamla (saadb@chbe.gatech.edu)

\section*{Acknowledgements}

This work was funded by NSF IRES \#2246236

\section*{DECLARATION OF INTERESTS}

Authors declare no competing interests.

\section*{SUPPLEMENTAL INFORMATION INDEX}




\begin{description}
  \item Supplemental 1: Sample gear checklist
  \item Supplemental 2: Participant survey

\end{description}

\newpage

\section*{MAIN FIGURE TITLES AND LEGENDS}




\noindent
\includegraphics[width=0.85\linewidth]{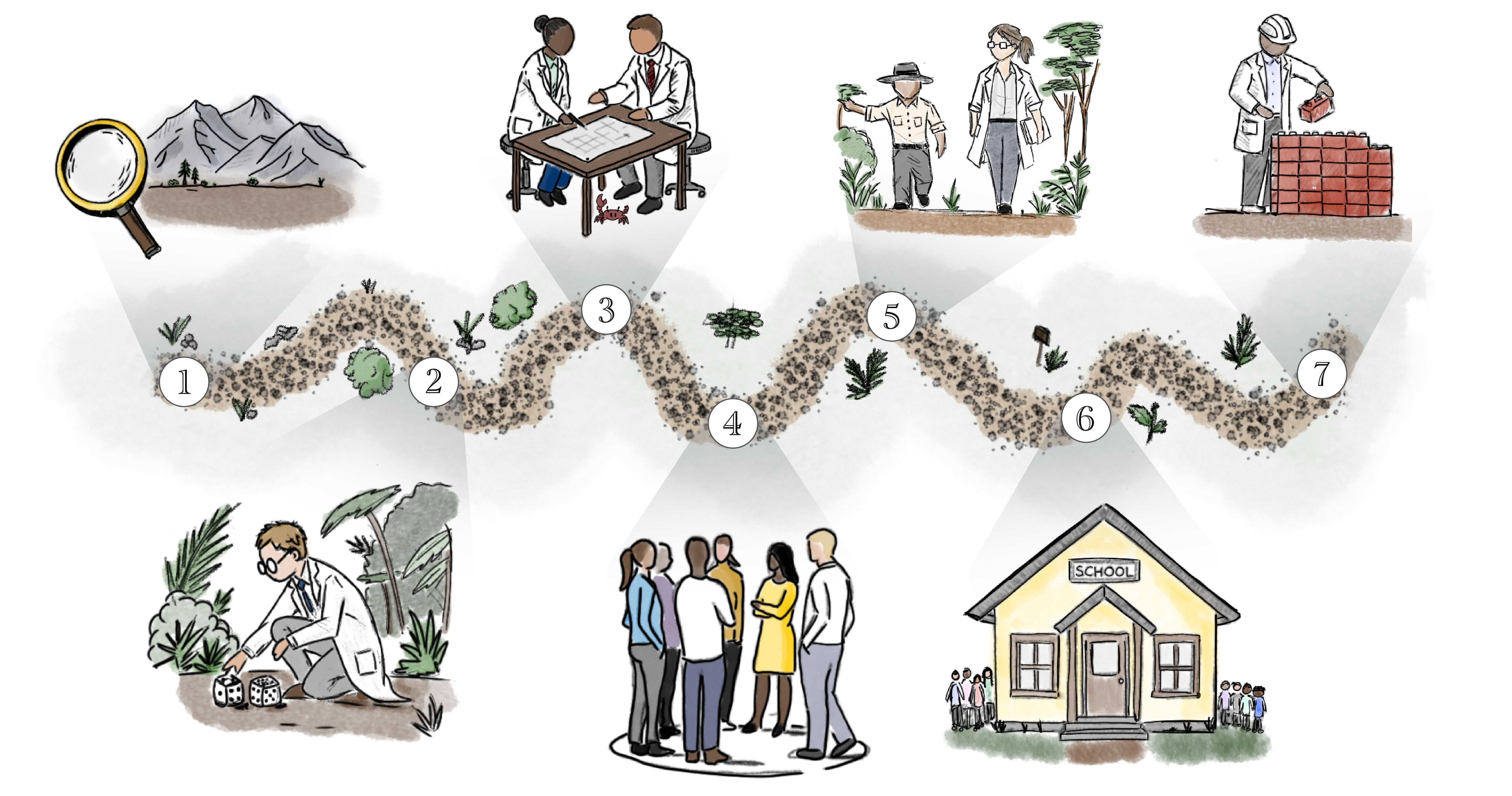}

\subsection*{Figure 1. Recommended steps for building an effective, collaborative, and safe field research program.} 

1. Find a field site for curiosity-driven research, 2. Identify new lines of research with emphasis on curiosity and observation, 3. Plan the trip meticulously and with a safety mindset, 4. Establish a culture of respect for all participants, 5. Collaborate with other scientists across disciplines, 6. Give back to local communities, 7. Iterate and Improve based the program based on feedback

\newpage

\section*{MAIN TABLES, INCLUDING TITLES AND LEGENDS}



\bibliography{references}

\begin{thebibliography}{22}
\providecommand{\natexlab}[1]{#1}
\providecommand{\url}[1]{\texttt{#1}}
\providecommand{\href}[2]{#2}
\providecommand{\path}[1]{#1}
\providecommand{\DOIprefix}{doi: }
\providecommand{\ArXivprefix}{arXiv: }
\providecommand{\URLprefix}{URL: }
\providecommand{\Pubmedprefix}{pmid: }
\providecommand{\doi}[1]{\href{http://dx.doi.org/#1}{\path{#1}}}
\providecommand{\Pubmed}[1]{\href{pmid:#1}{\path{#1}}}
\providecommand{\BIBand}{and}
\providecommand{\bibinfo}[2]{#2}
\ifx\xfnm\undefined \def\xfnm[#1]{\unskip,\space#1}\fi
\makeatletter\def\@biblabel#1{#1.}\makeatother
\bibitem[{Challita~J. et~al.(2023)Challita~J., Sehgal, Ko, Krugner and Bhamla~M}]{challitanaturecomm2023}
\bibinfo{author}{Challita~J., E.}, \bibinfo{author}{Sehgal, P.}, \bibinfo{author}{Ko, H.}, \bibinfo{author}{Krugner, R.}, and \bibinfo{author}{Bhamla~M, S.} (\bibinfo{year}{2023}). \bibinfo{title}{Droplet superpropulsion in an energetically constrained insect}.
\newblock \bibinfo{journal}{Nature Communications} \emph{\bibinfo{volume}{14}}.
\bibitem[{Sarfati et~al.(2021)Sarfati, Hayes and Peleg}]{sarfati2021self}
\bibinfo{author}{Sarfati, R.}, \bibinfo{author}{Hayes, J.C.}, and \bibinfo{author}{Peleg, O.} (\bibinfo{year}{2021}). \bibinfo{title}{Self-organization in natural swarms of photinus carolinus synchronous fireflies}.
\newblock \bibinfo{journal}{Science Advances} \emph{\bibinfo{volume}{7}}, \bibinfo{pages}{eabg9259}.
\bibitem[{England and Robert(2024)}]{england2024prey}
\bibinfo{author}{England, S.J.}, and \bibinfo{author}{Robert, D.} (\bibinfo{year}{2024}). \bibinfo{title}{Prey can detect predators via electroreception in air}.
\newblock \bibinfo{journal}{Proceedings of the National Academy of Sciences} \emph{\bibinfo{volume}{121}}, \bibinfo{pages}{e2322674121}.
\bibitem[{Guo et~al.(2023)Guo, Himsworth, Lee and Byers}]{guo2023systematic}
\bibinfo{author}{Guo, X.}, \bibinfo{author}{Himsworth, C.G.}, \bibinfo{author}{Lee, M.J.}, and \bibinfo{author}{Byers, K.A.} (\bibinfo{year}{2023}). \bibinfo{title}{A systematic review of rat ecology in urban sewer systems}.
\newblock \bibinfo{journal}{Urban Ecosystems} \emph{\bibinfo{volume}{26}}, \bibinfo{pages}{223--232}.
\bibitem[{Tydecks et~al.(2016)Tydecks, Bremerich, Jentschke, Likens and Tockner}]{tydecks2016biological}
\bibinfo{author}{Tydecks, L.}, \bibinfo{author}{Bremerich, V.}, \bibinfo{author}{Jentschke, I.}, \bibinfo{author}{Likens, G.E.}, and \bibinfo{author}{Tockner, K.} (\bibinfo{year}{2016}). \bibinfo{title}{Biological field stations: A global infrastructure for research, education, and public engagement}.
\newblock \bibinfo{journal}{BioScience} \emph{\bibinfo{volume}{66}}, \bibinfo{pages}{164--171}.
\bibitem[{noa(2020)}]{noauthor_undated-ft}
 (\bibinfo{year}{2020}).
\newblock \bibinfo{title}{{NSF} award search: Award \# 1941933 - {CAREER}: Fast, furious and fantastic beasts: Integrative principles, biomechanics and physical limits of impulsive motion in ultrafast organisms}.
\newblock \bibinfo{howpublished}{\url{https://www.nsf.gov/awardsearch/showAward?AWD_ID=1941933&HistoricalAwards=false}}. .
\newblock \URLprefix \url{https://www.nsf.gov/awardsearch/showAward?AWD_ID=1941933&HistoricalAwards=false} \bibinfo{note}{accessed: 2024-6-18}.
\bibitem[{noa(2023)}]{noauthor_undated-ci}
 (\bibinfo{year}{2023}).
\newblock \bibinfo{title}{{NSF} award search: Award \# 2246236 - {IRES} track1: In-situ jungle biomechanics laboratory ({JBL}) research experience in the amazon rainforest}.
\newblock \bibinfo{howpublished}{\url{https://www.nsf.gov/awardsearch/showAward?AWD_ID=2246236&HistoricalAwards=false}}. .
\newblock \URLprefix \url{https://www.nsf.gov/awardsearch/showAward?AWD_ID=2246236&HistoricalAwards=false} \bibinfo{note}{accessed: 2024-6-15}.
\bibitem[{Hunt et~al.(2021)Hunt, Jinn, Jacobs and Full}]{science2021hunt}
\bibinfo{author}{Hunt, N.}, \bibinfo{author}{Jinn, J.}, \bibinfo{author}{Jacobs, F.}, and \bibinfo{author}{Full, J.} (\bibinfo{year}{2021}). \bibinfo{title}{Acrobatic squirrels learn to leap and land on tree branches without falling}.
\newblock \bibinfo{journal}{Science} \emph{\bibinfo{volume}{373}}, \bibinfo{pages}{697--700}.
\bibitem[{Gart et~al.(2015)Gart, Socha, Vlachos and Jung}]{gart2015dogs}
\bibinfo{author}{Gart, S.}, \bibinfo{author}{Socha, J.J.}, \bibinfo{author}{Vlachos, P.P.}, and \bibinfo{author}{Jung, S.} (\bibinfo{year}{2015}). \bibinfo{title}{Dogs lap using acceleration-driven open pumping}.
\newblock \bibinfo{journal}{Proceedings of the National Academy of Sciences} \emph{\bibinfo{volume}{112}}, \bibinfo{pages}{15798--15802}.
\bibitem[{Ortega-Jimenez~M. et~al.(2022)Ortega-Jimenez~M., Challita~J., Kim, Ko, Gwon, Koh and Bhamla~M}]{PNAS2022ortegajimenez}
\bibinfo{author}{Ortega-Jimenez~M., V.}, \bibinfo{author}{Challita~J., E.}, \bibinfo{author}{Kim, B.}, \bibinfo{author}{Ko, H.}, \bibinfo{author}{Gwon, M.}, \bibinfo{author}{Koh, J.S.}, and \bibinfo{author}{Bhamla~M, S.} (\bibinfo{year}{2022}). \bibinfo{title}{Directional takeoff, aerial righting, and adhesion landing of semiaquatic springtails}.
\newblock \bibinfo{journal}{PNAS} \emph{\bibinfo{volume}{119}}, \bibinfo{pages}{e2211283119}.
\bibitem[{Flaum and Prakash(2024)}]{flaum2024curved}
\bibinfo{author}{Flaum, E.}, and \bibinfo{author}{Prakash, M.} (\bibinfo{year}{2024}). \bibinfo{title}{Curved crease origami and topological singularities enable hyperextensibility of l. olor}.
\newblock \bibinfo{journal}{Science} \emph{\bibinfo{volume}{384}}, \bibinfo{pages}{eadk5511}.
\bibitem[{Noel and Hu(2018)}]{noel2018cats}
\bibinfo{author}{Noel, A.C.}, and \bibinfo{author}{Hu, D.L.} (\bibinfo{year}{2018}). \bibinfo{title}{Cats use hollow papillae to wick saliva into fur}.
\newblock \bibinfo{journal}{Proceedings of the National Academy of Sciences} \emph{\bibinfo{volume}{115}}, \bibinfo{pages}{12377--12382}.
\bibitem[{Challita and Bhamla(2024)}]{challita2024unifying}
\bibinfo{author}{Challita, E.J.}, and \bibinfo{author}{Bhamla, M.S.} (\bibinfo{year}{2024}). \bibinfo{title}{Unifying fluidic excretion across life from cicadas to elephants}.
\newblock \bibinfo{journal}{Proceedings of the National Academy of Sciences} \emph{\bibinfo{volume}{121}}, \bibinfo{pages}{e2317878121}.
\bibitem[{Burrows and Sutton(2013)}]{burrows2013interacting}
\bibinfo{author}{Burrows, M.}, and \bibinfo{author}{Sutton, G.} (\bibinfo{year}{2013}). \bibinfo{title}{Interacting gears synchronize propulsive leg movements in a jumping insect}.
\newblock \bibinfo{journal}{science} \emph{\bibinfo{volume}{341}}, \bibinfo{pages}{1254--1256}.
\bibitem[{Patek et~al.(2004)Patek, Korff and Caldwell}]{patek2004deadly}
\bibinfo{author}{Patek, S.N.}, \bibinfo{author}{Korff, W.}, and \bibinfo{author}{Caldwell, R.L.} (\bibinfo{year}{2004}). \bibinfo{title}{Deadly strike mechanism of a mantis shrimp}.
\newblock \bibinfo{journal}{Nature} \emph{\bibinfo{volume}{428}}, \bibinfo{pages}{819--820}.
\bibitem[{Schulz et~al.(2022)Schulz, Boyle, Boyle, Sordilla, Rincon, Hooper, Aubuchon, Reidenberg, Higgins and Hu}]{schulz2022skin}
\bibinfo{author}{Schulz, A.K.}, \bibinfo{author}{Boyle, M.}, \bibinfo{author}{Boyle, C.}, \bibinfo{author}{Sordilla, S.}, \bibinfo{author}{Rincon, C.}, \bibinfo{author}{Hooper, S.}, \bibinfo{author}{Aubuchon, C.}, \bibinfo{author}{Reidenberg, J.S.}, \bibinfo{author}{Higgins, C.}, and \bibinfo{author}{Hu, D.L.} (\bibinfo{year}{2022}). \bibinfo{title}{Skin wrinkles and folds enable asymmetric stretch in the elephant trunk}.
\newblock \bibinfo{journal}{Proceedings of the National Academy of Sciences} \emph{\bibinfo{volume}{119}}, \bibinfo{pages}{e2122563119}.
\bibitem[{Catania(2016)}]{catania2016leaping}
\bibinfo{author}{Catania, K.C.} (\bibinfo{year}{2016}). \bibinfo{title}{Leaping eels electrify threats, supporting humboldt’s account of a battle with horses}.
\newblock \bibinfo{journal}{Proceedings of the National Academy of Sciences} \emph{\bibinfo{volume}{113}}, \bibinfo{pages}{6979--6984}.
\bibitem[{Whitesides(2018)}]{Whitesides2018-wb}
\bibinfo{author}{Whitesides, G.M.} (\bibinfo{year}{2018}). \bibinfo{title}{Curiosity and science}.
\newblock \bibinfo{journal}{Angewandte Chemie} \emph{\bibinfo{volume}{57}}, \bibinfo{pages}{4126--4129}. \URLprefix \url{http://dx.doi.org/10.1002/anie.201800684}. \DOIprefix\doi{10.1002/anie.201800684}.
\bibitem[{Demery and Pipkin(2021)}]{demery2021natureec}
\bibinfo{author}{Demery, A.J.C.}, and \bibinfo{author}{Pipkin, M.A.} (\bibinfo{year}{2021}). \bibinfo{title}{Safe fieldwork strategies for at-risk individuals, their supervisors and institutions}.
\newblock \bibinfo{journal}{Nature Ecology \& Evolution} \emph{\bibinfo{volume}{5}}, \bibinfo{pages}{5--9}.
\bibitem[{{Association of Polar Early Career Scientists (APECS)}(nd)}]{apecs_field_code_conduct}
\bibinfo{author}{{Association of Polar Early Career Scientists (APECS)}} (\bibinfo{year}{n.d.}).
\newblock \bibinfo{title}{{Field Code of Conduct}}.
\newblock \bibinfo{howpublished}{\url{https://www.apecs.is/career-resources/diversity-equity-inclusion/field-code-of-conduct.html}}. .
\newblock \bibinfo{note}{Accessed: 2025-06-23}.
\bibitem[{Odeny and Bosurgi(2022)}]{odeny2022time}
\bibinfo{author}{Odeny, B.}, and \bibinfo{author}{Bosurgi, R.} (\bibinfo{year}{2022}).
\newblock \bibinfo{title}{Time to end parachute science}. \bibinfo{publisher}{Public Library of Science San Francisco, CA USA}.
\bibitem[{Clayton and Opotow(2003)}]{clayton2003}
\bibinfo{author}{Clayton, S.}, and \bibinfo{author}{Opotow, S.} (\bibinfo{year}{2003}). \bibinfo{title}{Identity and the natural environment: The psychological significance of nature}. \bibinfo{publisher}{MIT Press}.

\end{thebibliography}

\bigskip


\newpage

\end{document}